\documentstyle[12pt]{article}
\begin{document}
\def\thefootnote{\fnsymbol{footnote}}
\begin{flushright}
KANAZAWA-05-02  \\ 
February, 2005
\end{flushright}
\vspace*{2cm}
\begin{center}
{\LARGE\bf Neutrino texture saturating the CP asymmetry}\\
\vspace{1 cm}
{\Large Takayasu Baba and}
{\Large Daijiro Suematsu}
\footnote[2]{e-mail:~suematsu@hep.s.kanazawa-u.ac.jp}
\vspace {1cm}\\
{\it Institute for Theoretical Physics, Kanazawa University,\\
        Kanazawa 920-1192, Japan}\\    
\end{center}
\vspace{1cm}
{\Large\bf Abstract}\\  
We study a neutrino mass texture which can explain the neutrino
oscillation data and also saturate the upper bound of the 
CP asymmetry $\varepsilon_1$ in the leptogenesis. We consider the 
thermal and non-thermal leptogenesis based on the right-handed 
neutrino decay in this model. 
A lower bound of the reheating temperature required for 
the explanation of the baryon number asymmetry is estimated 
as $O(10^8)$GeV for the thermal 
leptogenesis and $O(10^{6})$GeV for the non-thermal one.
It can be lower than the upper bound of the reheating temperature 
imposed by the cosmological gravitino problem.
An example of the construction of the discussed texture is 
also presented.
\newpage
\setcounter{footnote}{0}
\def\thefootnote{\arabic{footnote}}
\section{Introduction}
The discovery of the neutrino masses \cite{sk} gives large 
impact to the study of particle physics and astroparticle physics. 
In particular, it presents an interesting approach to the study 
of the origin of the baryon number ($B$) asymmetry in the universe, 
which is one of the important questions in these fields. 
The leptogenesis \cite{lept} based on the $B-L$ violation due to the 
neutrino masses is considered to be the most promising scenario for 
the generation of the $B$ asymmetry. 
During the recent few years, the leptogenesis based on 
the CP asymmetric decay of the heavy right-handed 
neutrinos \cite{cpasym} whose existence is required by the seesaw mechanism 
\cite{seesaw} has been extensively studied \cite{leptg0,leptg1}.

Since the intermediate scale is generally necessary for 
the seesaw mechanism, it seems to be natural to consider the
leptogenesis in the supersymmetric framework to guarantee the 
stability of that scale against the radiative correction.
However, if we consider it in such a framework,
a crucial problem called the cosmological gravitino problem 
is caused in relation to the generation of the right-handed neutrinos. 
If the reheating temperature $T_R$ required to produce a sufficient
amount of the right-handed neutrinos is a high value such as
$T_R~{^>_\sim}~10^8$GeV, the gravitino can be produced too 
much and its late time decay may disturb the nucleosynthesis 
\cite{nucl}.\footnote{If the gravitino is
the lightest superparticle as in the gauge mediated
supersymmetry breaking, there is no gravitino problem even in the case
of $T_R=O(10^{10})$GeV \cite{solgmsb}.}
Thus, the production mechanism of the heavy right-handed neutrinos 
is the important ingredient for this problem.
Several solutions for this difficulty have been proposed by now
\cite{inonth,fnonth}.

On the other hand, the CP asymmetry \cite{cpasym} in the decay of 
the right-handed neutrinos is another crucial factor which plays the 
essential role to determine the generated lepton number ($L$) asymmetry. 
Its magnitude depends on the structure of the neutrino mass matrix,
which is severely constrained to 
explain the neutrino oscillation data \cite{sk}.
From a viewpoint of the leptogenesis, it is favorable that the neutrino
mass matrix can realize the maximum value of 
the CP asymmetry \cite{cpasym,di} or enhance its value \cite{dege,soft}.  
Thus it is important for the quantitative study of
the leptogenesis to construct such a concrete model 
for the neutrino mass matrix as done in \cite{leptg0,by,leptg2}
and to proceed the investigation based on it. 
In this paper we present an example of the neutrino mass matrix
and estimate the reheating temperature required to produce 
the sufficient $B$ asymmetry. 

The paper is organized as follows. In section 2 we present a neutrino
mass texture and discuss its phenomenological features. An example for
its construction is discussed in appendix A. In section 3 we apply 
this model to both the thermal and non-thermal leptogenesis. 
We discuss the lower bound of the reheating temperature 
required for the production of the sufficient $B$ asymmetry. 
Section 4 is devoted to the summary.

\section{Neutrino mass texture}
We consider the minimal supersymmetric standard model (MSSM) extended
with gauge singlet chiral superfields $N_i$ which correspond to 
three generation right-handed neutrinos. 
An effective superpotential for the neutrino sector is assumed as follows: 
\begin{equation}
W=\sum_{i,j=1}^3\left(h^\nu_{ij} N_i H_2 L_j 
+{1\over 2}{\cal M}_{ij} N_iN_j\right),
\label{superp}
\end{equation}
where $L_i$ and $H_2$ are the lepton doublet and Higgs doublet chiral
superfields, respectively. In this paper we use the same notation 
for both a superfield and its component fields. 
The right-handed neutrino mass matrix ${\cal M}$ and the Dirac mass matrix
$m_D$ induced from the first term in this superpotential 
are assumed to take the form as\footnote{This model can be 
considered as a simple extension of the model in \cite{fgy}.
It is realized by adding a right-handed neutrino $N_1$ to the original
one in such a way that $N_1$ weakly couples to $N_2$ alone. }
\begin{equation}
{\cal M}\equiv\left(\begin{array}{ccc}
M_1 & m & 0 \\ m & M_2 & 0 \\ 0 & 0& M_3\\ 
\end{array}\right), \qquad
m_D\equiv h^\nu\langle H_2\rangle=\left(\begin{array}{ccc}
0 & 0& 0\\ a & a^\prime & 0 \\ 0 & b & b^\prime \\
\end{array}\right),
\label{mmatrix}
\end{equation} 
where the charged lepton mass matrix is considered to be diagonal. 
Although each element of ${\cal M}$ is supposed to be real, $m_D$ 
is assumed to be a complex matrix.

If the hierarchical structure 
\begin{equation}
m, ~M_1\ll M_2 \ll M_3 
\end{equation}
is assumed in the right-handed neutrino sector, the eigenvalues
$\tilde M_i$ of the mass matrix ${\cal M}$ can be approximated to 
\begin{equation}
\begin{array}{llccl}
{\rm (a)}~ & \tilde M_1(\simeq M_1), &  M_2, &  M_3, & \quad 
({\rm for}~m^2 <M_1M_2), \\   
{\rm (b)}~ &\tilde M_1(\simeq M_2\sin^2\xi),&  
M_2, &  M_3, & \quad ({\rm for}~m^2 >M_1M_2), \\  
\end{array}
\label{hier}
\end{equation}
where we use $\tilde M_i\simeq M_i~(i=2,3)$ and $\sin\xi\simeq m/M_2$.
These two cases are studied in the following part.
The structure of ${\cal M}$ and $m_D$ can be effectively realized 
by imposing a suitable symmetry on the superpotential at the high energy
scales. 
We give such an example for the construction of ${\cal M}$ and
$m_D$ in appendix A.  

If we change $N_i$ into the ${\cal M}$ diagonal basis $\tilde N_i$,
the Dirac neutrino mass matrix is transformed into
\begin{equation}
\tilde m_D\equiv \tilde h^\nu \langle H_2\rangle
=\left(\begin{array}{ccc}
-a\sin\xi & -a^\prime\sin\xi & 0\\
a\cos\xi & a^\prime\cos\xi & 0 \\
0 & b & b^\prime \\ \end{array} \right).
\label{dirac}
\end{equation}
Applying the seesaw mechanism to these matrices, we can obtain the
light neutrino mass eigenvalues and the MNS matrix.
Here, for the simplicity, we put $a^\prime=\sqrt 2a$ and 
$b^\prime=b$,\footnote{We adopt these relations in the following study.
Under this assumption the model contains seven real parameters.
We can loose these strict equalities without changing largely 
the qualitative results given in this paper.}  
and then the light neutrino mass eigenvalues are found to be
\begin{equation}
m_1=0, \qquad m_2\simeq {2|a|^2\over M_2}, \qquad 
m_3\simeq {2|b|^2 \over M_3}.
\label{mass}
\end{equation}
The MNS matrix has the bi-large mixing form such as 
\begin{equation}
U\simeq\left(\begin{array}{ccc}
{1\over\sqrt 2} & {1\over\sqrt 2} & {1\over\sqrt 2}\sin\theta \\
-{1\over 2} &{1\over 2}\cos\theta & {1\over\sqrt 2}\cos\theta \\
{1\over 2} & -{1\over 2}\cos\theta &{1\over\sqrt 2}\cos\theta \\
\end{array}\right),
\end{equation}
where we neglect the CP phase in this expression.

Now we can compare these results with the present experimental data. 
Since the neutrino mass eigenvalues are assumed to be hierarchical, 
the analysis for the neutrino oscillation experiments requires \cite{sk}
\begin{equation}
{2|a|^2\over M_2}\simeq \sqrt{\Delta m_{\rm sol}^2}
\simeq (7\times 10^{-5}~{\rm eV}^2)^{1/2}, \quad
{2|b|^2\over M_3}\simeq \sqrt{\Delta m_{\rm atm}^2}
\simeq (2\times 10^{-3}~{\rm eV}^2)^{1/2}, 
\label{nmass}
\end{equation}
and $\sin\theta$ should satisfy
\begin{equation}
\sin\theta\simeq{\vert a\vert^2/M_2 \over\sqrt 2\vert b\vert^2/M_3}\simeq 
{1\over\sqrt 2}\sqrt{\Delta m^2_{\rm sol}\over \Delta m^2_{\rm atm}}\sim 0.1. 
\end{equation}
This is consistent with the constraint $\sin\theta<0.16$ which is imposed
by the CHOOZ experiment \cite{chooz}.
The effective mass for the neutrinoless double $\beta$-decay is
estimated in this model as
\begin{equation}
m_{ee}~{^<_\sim}~{1\over 2}\left| \sqrt{\Delta m_{\rm atm}^2}\sin^2\theta
+\sqrt{\Delta m_{\rm sol}^2}\cos^2\theta\right|\sim 2\times 10^{-3}.
\end{equation}
It seems to be difficult to reach such a value in the next generation 
experiment.

The lepton flavor violating processes such as $\mu\rightarrow e\gamma$
can constrain the model. It has been suggested that these processes
could impose the strong constraint because of the renormalization effect
on the soft supersymmetry (SUSY) breaking parameters due 
to the off-diagonal Yukawa couplings. 
It can be very severe even in the case of the universal SUSY breaking 
in the gravity mediation scenario \cite{mueg}. 
Here in order to find the conservative condition, we consider the 
universal soft SUSY breaking in the gravity mediation.
The branching ratio of the flavor changing process 
$\ell_i\rightarrow\ell_j\gamma$ is estimated by
taking account of the one-loop contribution as \cite{mueg} 
\begin{equation}
Br(\ell_i\rightarrow\ell_j\gamma)=
{\alpha^3\tan^2\beta\over G_F^2m_{\tilde\ell}^8}
\left| {-1\over 8\pi^2}(3m_0^2+A_0^2)(\tilde h^{\nu\dagger}\tilde h^\nu)_{ij}
\ln{M_X\over M}\right|^2,
\end{equation}
where $m_0$, $A_0$ and $m_{\tilde\ell}$ represent the soft scalar mass, 
the SUSY breaking $A$ parameter and the relevant slepton mass,
respectively. $M_X$ stands for the unification scale 
and $M$ is the right-handed neutrino mass scale. 
Since $\tilde M_1$ is irrelevant to the light neutrino
masses as shown in eq.~(\ref{mass}), 
$M$ is appropriate to be taken as $M_2$ in the present model. 

If we assume $m_0\simeq m_{\tilde\ell}\simeq A_0$ and use
eq.~(\ref{dirac}) for the Yukawa couplings $h^\nu$, each branching ratio 
for $\mu\rightarrow e\gamma$ and $\tau\rightarrow\mu\gamma$ is estimated 
as\footnote{The decay $\tau\rightarrow e\gamma$ is automatically 
forbidden as a result of the present texture of the neutrino mass matrix.}
\begin{eqnarray}
&&Br(\mu\rightarrow e\gamma)\simeq 3\times 10^{-31}{M_2^2\over m_0^4}
\left(\ln{M_X\over M_2}\right)^2\tan^2\beta\le
1.2\times 10^{-11}, \nonumber\\
&&Br(\tau\rightarrow \mu\gamma)\simeq 4\times 10^{-30}{M_3^2\over m_0^4}
\left(\ln{M_X\over M_2}\right)^2\tan^2\beta \le
1.1\times 10^{-6}.
\end{eqnarray}
If we take $M_X=10^{16}$~GeV and $m_0=100$~GeV, for example, 
the present experimental bounds can be satisfied for
$M_2~{^<_\sim}~10^{11}$GeV and $M_3~{^<_\sim}~10^{13}$~GeV even 
in the case of large $\tan\beta$ such as $\tan\beta\simeq 50$.   
This means that no lepton flavor violating decays 
$\ell_i\rightarrow \ell_j\gamma$ contradict with the present
 model as far as the universal SUSY breaking is assumed even in the
 gravity mediation scenario. 

A remarkable feature of the model is that there are no constraints 
on $M_1$ and $\sin\xi$ from the neutrino oscillation data and 
other present available experiments. 
If we apply this neutrino mass texture to the leptogenesis, 
 $M_1$ and $\sin\xi$ may be constrained to explain the $B$ asymmetry. 
In the next section we focus our study on this point. 

\section{Application to leptogenesis}
The decay of the heavy right-handed neutrinos can produce the $B-L$ 
asymmetry. Then it may explain the $B$ asymmetry in the universe 
since the sphaleron interaction can convert a part of the $B-L$ asymmetry 
into the $B$ asymmetry \cite{sph}.
The $L$ asymmetry or the $B-L$ asymmetry induced through 
this decay is produced as a result of the CP asymmetry caused by the 
interference between the tree and one-loop diagrams.

The CP asymmetry appeared in the $\tilde N_i$ decay can be generally 
expressed as \cite{cpasym} 
\begin{eqnarray}
\varepsilon_i&\equiv&
{\sum_j\Gamma(\tilde N_i\rightarrow L_jH_2)-
\sum_j\Gamma(\tilde N_i\rightarrow \bar L_j\bar H_2)\over
\sum_j\Gamma(\tilde N_i\rightarrow L_jH_2)
+\sum_j\Gamma(\tilde N_i\rightarrow \bar L_j\bar H_2)} \nonumber\\
&=&-{1\over 8\pi}
{1\over(\tilde h^\nu \tilde h^{\nu\dagger})_{ii}}\sum_{k\not=i}
{\rm Im}[(\tilde h^\nu \tilde h^{\nu\dagger})^2_{ik}]~f
\left({M_k^2\over M_i^2}\right),
\label{leptpara}
\end{eqnarray} 
where $f(x)$ contains the contributions from both the vertex correction 
and the self-energy correction, and it has an expression
\begin{equation}
f(x)=\sqrt{x}\left[\ln\left({1+x\over x}\right)+{2\over x-1}\right].
\end{equation}
Applying this formula to our model in which the hierarchical structure
for the right-handed neutrino masses is assumed, we obtain 
\begin{eqnarray}
&&\varepsilon_1\simeq{1\over 8\pi}
{\sqrt{\Delta m_{\rm atm}^2}\tilde M_1\over v^2\sin^2\beta}\sin2\chi, \qquad
\varepsilon_2\simeq{1\over 8\pi}
{\sqrt{\Delta m_{\rm atm}^2} M_2\over v^2\sin^2\beta}\sin2\chi, \nonumber\\
&&\varepsilon_3\simeq-{1\over 16\pi}
{\sqrt{\Delta m_{\rm sol}^2}M_2\over v^2\sin^2\beta}
\left({\tilde M_1\over M_3}\ln{M_3\over \tilde M_1}\sin^2\xi+
{M_2\over M_3}\ln{M_3\over M_2}\cos^2\xi\right)\sin2\chi,
\label{asym}
\end{eqnarray} 
where $\langle H_2\rangle\equiv v\sin\beta$ and $\chi\equiv \arg(a^\ast b)$.
The formulas in eq.~(\ref{asym}) show that $\varepsilon_3$ can be much smaller 
than $\varepsilon_{1,2}$. 
The interesting point of this result is that $\varepsilon_1$ is 
almost equal to the expression which saturates its upper bound given 
in \cite{di}.\footnote{
A model with this feature has been discussed in \cite{by} already. 
However, it seems not to have a satisfactory structure for the explanation of
the neutrino oscillation data.}
Thus the present texture for the neutrino masses seems to be favorable to
induce the $B$ asymmetry.
 
By using these expression for $\varepsilon_i$, the $L$ asymmetry resulting from
the decay of the right-handed neutrinos can be estimated.
If we put the excess of the number density of the right-handed neutrino
$\tilde N_i$ from the equilibrium one as $n_i$
and the entropy density in the comoving volume 
at the latest $\tilde N_i$ decay as $s$, the produced $L$ asymmetry 
through this decay can be expressed as
\begin{equation}
Y_L\equiv {n_L\over s}\simeq \sum_{i=1}^3{2n_i\over s}\varepsilon_i\kappa_i,
\label{leptasym}
\end{equation}
where $\kappa_i$ represents the washout effect which depends on the
strength of the Yukawa couplings $\tilde h^\nu_{ij}$ in eq.~(\ref{dirac}). 
The sphaleron interaction which is in the thermal equilibrium at 
the temperature $10^2~{\rm GeV}~{^<_\sim}~
T_{\rm sph}~{^<_\sim}~10^{12}~{\rm GeV}$
converts a part of the $B-L$ asymmetry into 
the $B$ asymmetry in such a way as $n_B/s=-(8/15)(n_L/s)$ 
in the MSSM case \cite{sph,susysph}. 
Thus $n_L/s$ should satisfy $|n_L/s| ~{^>_\sim}~ 10^{-10}$ to realize the 
observed value $n_B/s\simeq (0.6-1)\times 10^{-10}$.

Since the number density $n_i$ of the right-handed neutrino $\tilde N_i$
depends on its generation mechanism, we need to fix it for the
quantitative estimation of $n_L/s$.
In the following study we consider both the thermal and non-thermal
scenarios for their generation. In the non-thermal leptogenesis 
we mainly consider that the right-handed neutrinos couple to 
the inflaton directly and then the inflaton 
decay produces the right-handed neutrinos non-thermally.   

\subsection{Thermal leptogenesis}
Since the right-handed neutrino masses are supposed to be hierarchical in the
present model, the $L$ asymmetry is expected to be produced as a result of the
out of equilibrium decay of the lightest right-handed neutrino $\tilde N_1$
as studied in a lot of works \cite{leptg0,leptg1}. 
In fact, it can be easily checked in the present model.
If we use eqs.~(\ref{dirac}) and (\ref{nmass}), 
the decay width of $\tilde N_i$ can be estimated as
\begin{eqnarray}
&&\Gamma_{\tilde N_1}\simeq{3\over 16\pi}{\sqrt{\Delta m_{\rm sol}^2}
\tilde M_1M_2\over v^2\sin^2\beta}\sin^2\xi
\sim\left( {\tilde M_1\over 10^7~{\rm GeV}}\right)
\left({M_2\over 10^{10}~{\rm GeV}}\right)\sin^2\xi, \nonumber \\
&&\Gamma_{\tilde N_2}\simeq{3\over 16\pi}{\sqrt{\Delta m_{\rm sol}^2}
M_2^2\over v^2\sin^2\beta}\cos^2\xi
\sim 10^3\left({M_2\over 10^{10}~{\rm GeV}}\right)^2\cos^2\xi, 
\nonumber \\
&&\Gamma_{\tilde N_3}\simeq{1\over 8\pi}{\sqrt{\Delta m_{\rm atm}^2}
M_3^2\over v^2\sin^2\beta}
\sim 10^9\left({M_3\over 10^{13}~{\rm GeV}}\right)^2, 
\label{decay}
\end{eqnarray}
where these decay widths are given in the GeV unit.
This shows that $\Gamma_{\tilde N_1}<\Gamma_{\tilde N_2}
<\Gamma_{\tilde N_3}$ is satisfied.
The asymmetry generated by the decay of $\tilde N_{2,3}$ is washed out 
through the $L$ violating scattering mediated by the 
thermal $\tilde N_1$ {\it etc.} and then $\kappa_{2,3}\ll 1$. 

If we use the thermal number density of the relativistic 
particle for $n_1$ and $s={2\pi^2\over45}g_\ast T^3 $ in
eq.(\ref{leptasym}), the $L$ asymmetry generated through 
the $\tilde N_1$ decay is expressed as
\begin{equation}
{n_L\over s}\simeq {1\over g_\ast}\varepsilon_1\kappa_1,
\end{equation}  
where $g_\ast$ is a degree of freedom for the relativistic particles at
this period and $g_\ast \sim 200$ in the MSSM. 
In this expression we should note
that $\kappa_1$ includes also the efficiency factor to generate $\tilde N_1$ 
in the thermal bath other than the washout effect since we suppose that there
are no thermal right-handed neutrinos initially. 

As is well known in the thermal leptogenesis \cite{leptg1}, 
there is an important quantity $\tilde m_1$ which is related to 
$\kappa_1$ and then the strength of the relevant Yukawa couplings of
$\tilde N_1$. 
It controls how many $\tilde N_1$ is produced in the thermal equilibrium 
and also how much $L$ asymmetry is washed out.
In the present model $\tilde m_1$ is expressed as
\begin{equation}
\tilde m_1\equiv (\tilde h^\nu \tilde h^{\nu\dagger})_{11}
{v^2\sin^2\beta\over \tilde M_1} 
=\left\{\begin{array}{lc}
\displaystyle {3\over 2}\sqrt{\Delta m_{\rm sol}^2}{M_2\over M_1}\sin^2\xi & 
\quad {\rm for~ (a)},\\
\displaystyle {3\over 2}\sqrt{\Delta m_{\rm sol}^2} &
\quad  {\rm for~ (b)}.\\
\end{array}\right.
\label{meff}
\end{equation}
As is found from eqs.~(\ref{asym}) and (\ref{meff}), 
$\varepsilon_1$ and $\tilde m_1$ can be independent from each other
because of the freedom of $\sin\xi$.  
We may expect that there is generally some 
correlation between these parameters from their definitions 
(\ref{leptpara}) and (\ref{meff}). 
However, the special texture can make them independent in the present model. 
This feature may cause a substantial influence on the reheating
temperature required for the leptogenesis.

If we use the formula in eq.~(\ref{asym}), the CP asymmetry $\varepsilon_1$ 
required from the $B$ asymmetry in the universe is estimated as 
\begin{equation}
\vert\varepsilon_1\vert\simeq 
10^{-8}\times\left({\tilde M_1\over 10^{8}~{\rm GeV}}\right)~
{^>_\sim}~ 10^{-8},
\label{mbound}
\end{equation} 
where $g_\ast \sim 200$ is used. In this estimation the maximum CP phase 
$\vert\sin 2\chi\vert\sim 1$ and $\sin\beta=1$ are also 
assumed.\footnote{This assumption for $\sin\beta$ brings no crucial difference
as far as $\tan\beta$ is in the interesting region such as
$1~{^<_\sim}~\tan\beta~{^<_\sim}~50$.} 
From this condition we obtain $\tilde M_1~{^>_\sim}~10^8$~GeV for the lower 
bound of the $\tilde N_1$ mass, which is the ordinary 
result in the out of equilibrium decay of the thermally
produced $\tilde N_1$.
On the other hand, the effective mass $\tilde m_1$ is estimated as
\begin{equation}
{\tilde m_1\over 10^{-2}~{\rm eV}}\simeq 
\left\{\begin{array}{lc}\displaystyle 
{M_2\over M_1}\sin^2\xi & \quad {\rm for~ (a)}, \\
1 &\quad {\rm for~(b)}. \\
\end{array} \right.
\label{eff}
\end{equation}
While $\tilde m_1$ takes a fixed value in the case (b), there is the
freedom $\sin\xi$ to tune $\tilde m_1$ into the desirable value 
in the case (a).
Eq.~(\ref{eff}) shows that the efficiency factor can be in the 
favorable region for the leptogenesis in the 
case (a) but it seems to be larger than the favorable value 
in the case (b) \cite{leptg0,leptg1,leptg2}.
 
The necessary condition for the out of equilibrium decay of $\tilde N_1$ is 
given by $H>\Gamma_{\tilde N_1}$.
If we use eq.(\ref{decay}) for this condition, the successful leptogenesis 
requires that the temperature $T$ at the period of the $\tilde N_1$ decay 
should satisfy 
\begin{equation}
T > T_{\rm min} \simeq \sqrt{\tilde M_1M_2}\sin\xi.
\label{temp}
\end{equation} 
In the case (b) we find $T_{\rm min}\simeq \tilde
M_1$ and then $T_R~{^>_\sim}~\tilde M_1$ should be satisfied.
Thus we expect the similar result for the efficiency factor 
to the previous works \cite{leptg0,leptg1}.
On the other hand, since the case (a) is realized for $M_1 > M_2\sin^2\xi$, 
we find that $T <\tilde M_1$ could be consistent with the condition for the 
out of equilibrium decay.
Such a situation seems to be realized for a 
sufficiently small $\sin\xi$ without 
conflicting with the neutrino oscillation data.
The small $\sin\xi$ can also make the washout effect negligible.
Although this seems to suggest the possibility that the reheating temperature 
$T_R$ may not be necessary to be high enough compared with $\tilde M_1$,
however, the sufficient number of $\tilde N_1$ may not be produced due to the 
Boltzmann suppression in the thermal equilibrium. 
We need to solve a set of Boltzmann equations numerically for the
quantitative study of the relation between $T_R$ and $\tilde M_1$.
\input epsf 
\begin{figure}[tb]
\begin{center}
\epsfxsize=9cm
\leavevmode
\epsfbox{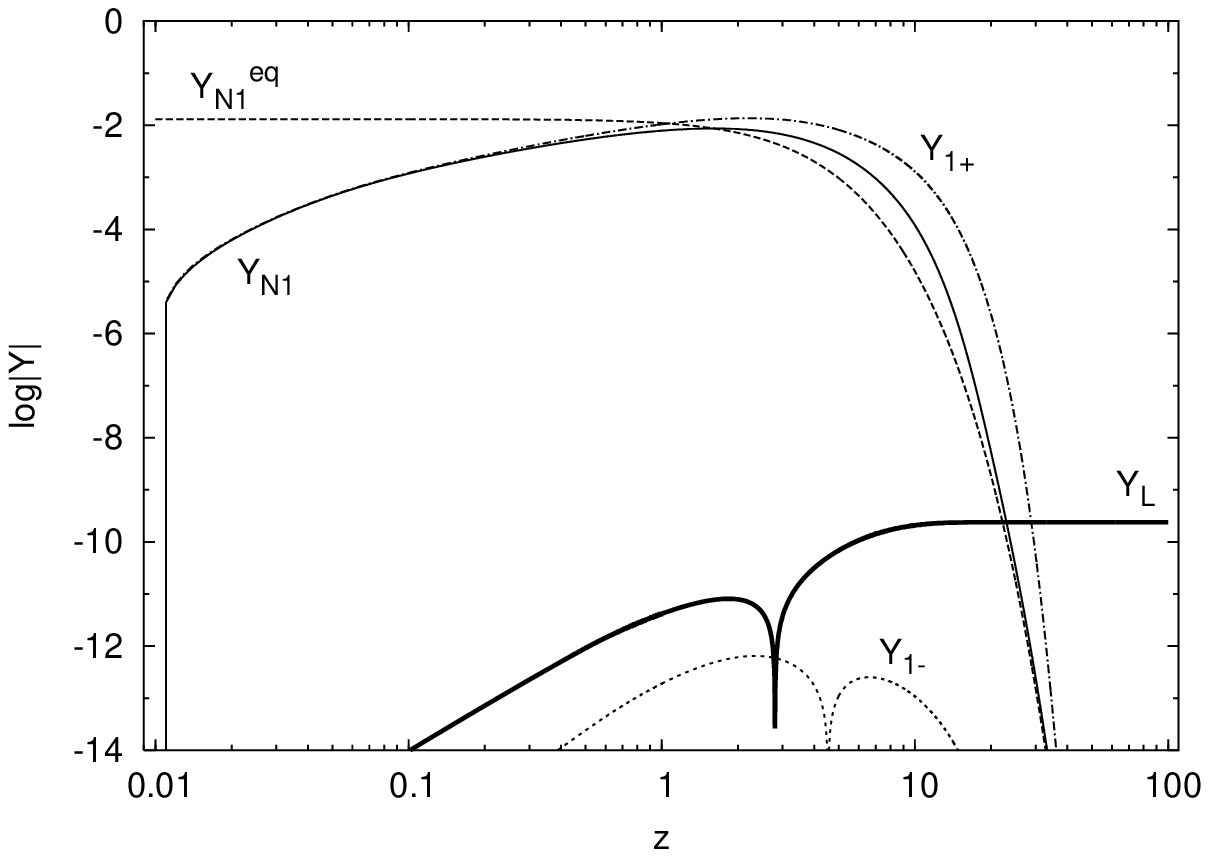}\\
\end{center}
\vspace*{-6mm}
{\footnotesize Fig. 1~~ A typical solution of the Boltzmann equations in the
 case of the thermal generation of $\tilde N_1$. We define $Y_i$ as 
$Y_i\equiv n_i/s~(i=\tilde N_1,\tilde N_2,L)$ and 
$Y_{\pm 1}\equiv(n_{SN_1}\pm \bar n_{SN_1})/s$ where $SN_i$ stands for the
 sneutrinos. $Y^{\rm eq}_{N_i}$ is the value in the equilibrium.}
\end{figure}

\begin{figure}[tb]
\begin{center}
\epsfxsize=7cm
\leavevmode
\epsfbox{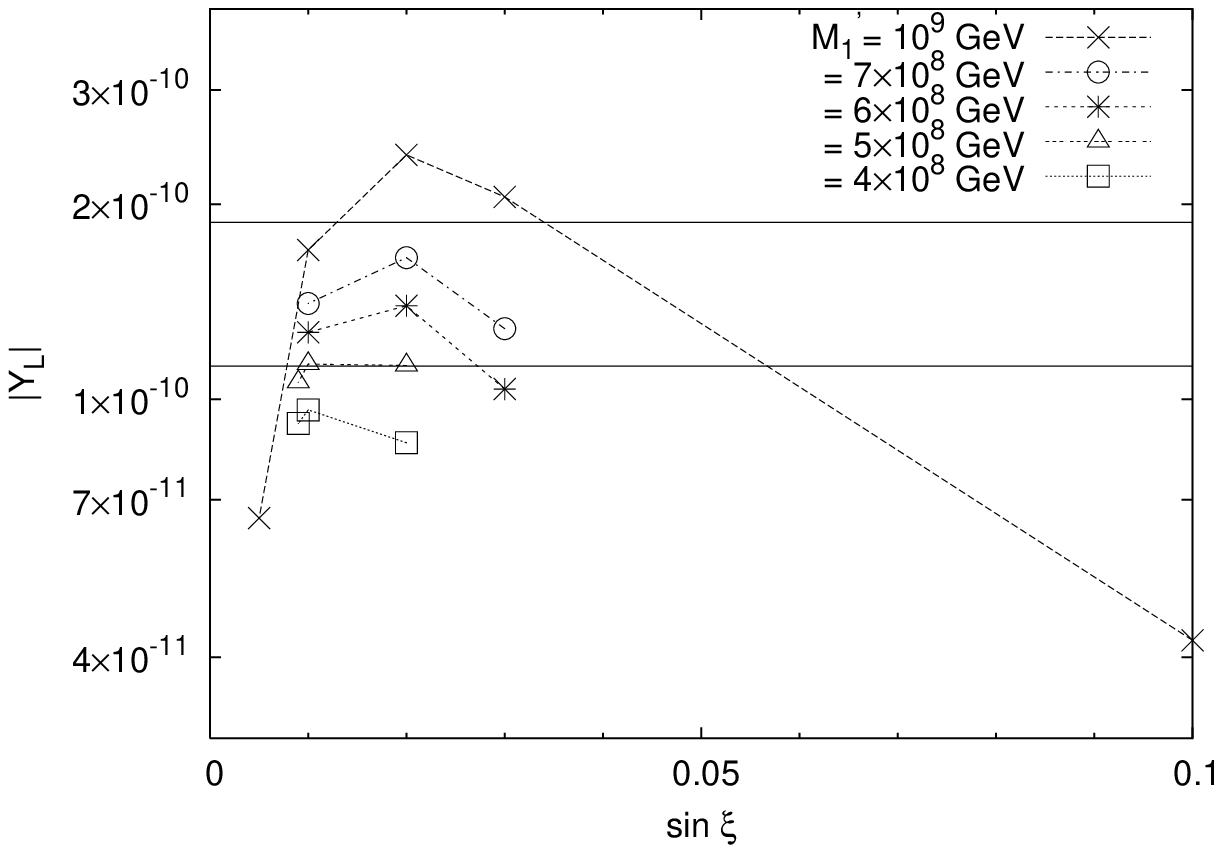}
\hspace*{7mm}
\epsfxsize=7cm
\leavevmode
\epsfbox{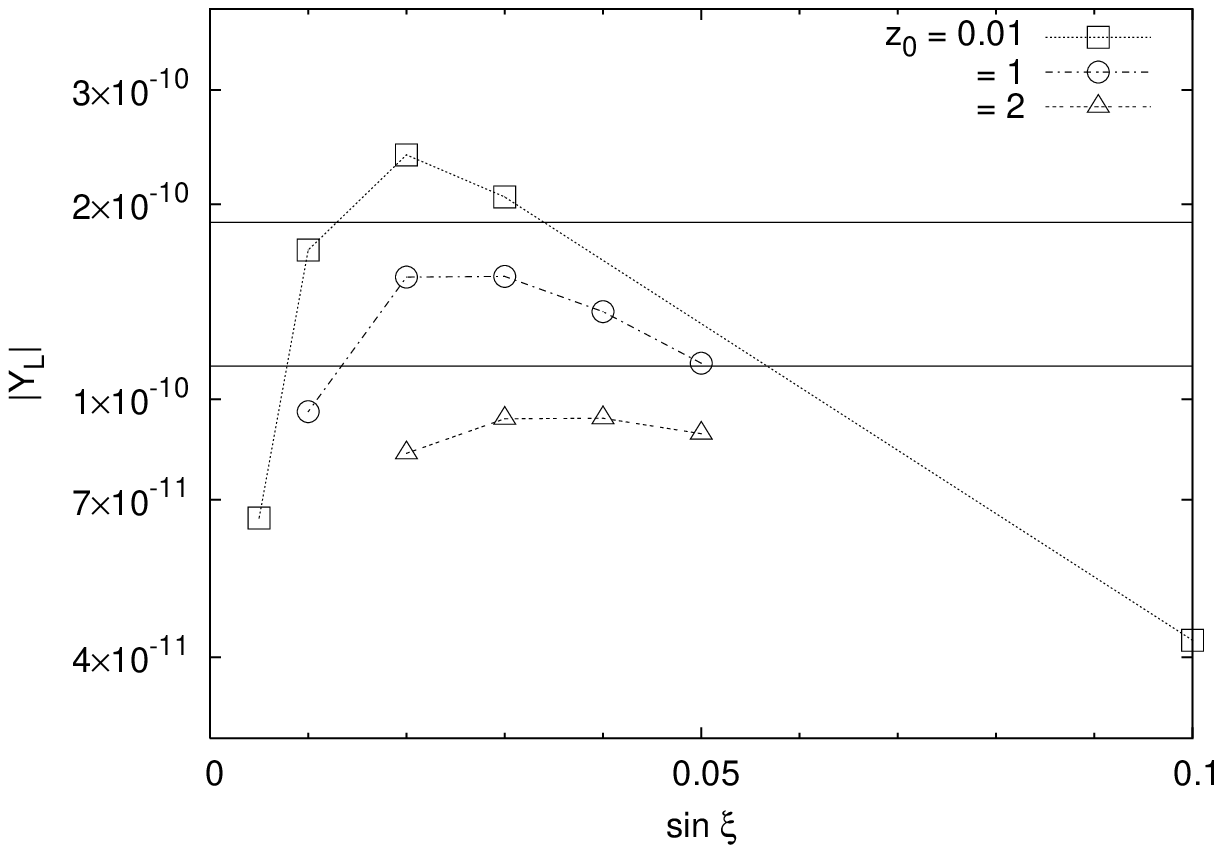}\\
\end{center}
\vspace*{-6mm}
{\footnotesize Fig. 2~~The $L$ asymmetry $|Y_L|$ as a function of
 $\sin\xi$. Horizontal thin lines represent the desirable region
 to explain the observed $B$ asymmetry. In the left panel $M_1$ is
 varied keeping other parameters fixed in such a way as $M_2=10^{10}$GeV, 
$M_3=10^{13}$GeV and $z_0=0.01$. In the right panel we vary the $z_0$
 value keeping others fixed as $M_1=10^9$GeV, $M_2=10^{10}$GeV and 
$M_3=10^{13}$GeV.}
\end{figure}
 
We study these points by solving numerically a set of Boltzmann
equations for the MSSM presented in \cite{leptg0}.
If we use eq.~(\ref{nmass}) and assume $|\sin 2\chi|=1$,
the model parameters in this calculation are $M_{1,2,3}$ and $\sin\xi$.
As an initial condition for the Boltzmann equations, we assume
that both the number density of $\tilde N_i$ and the $L$ asymmetry are
zero at $z_0=0$.
A dimensionless parameter $z$ is defined as $z=\tilde M_1/T$.
The temperature corresponding to $z_0$ may be considered to correspond to
 the reheating temperature $T_R$.

In Fig.~1 we give a solution of the Boltzmann equations with
$z_0=0.01$  and the input parameters such as
$M_1=10^9$GeV, $M_2=10^{10}$GeV, $M_3=10^{13}$GeV and $\sin\xi=0.02$.
This corresponds to the case (a) with $T_R=10^{11}$GeV.
The case (b) cannot yield the sufficient $L$ asymmetry. 
This can be understood as follows. 
In this case $\sin\xi$ is required to take a rather large value to realize 
$M_1<M_2\sin^2\xi$ satisfying both constraints
such as $\tilde M_1~{^>_\sim}~10^8$GeV and $M_3~{^<_\sim}~10^{13}$GeV,
which are imposed by the previously discussed phenomenological constraints. 
Such a $\sin\xi$ makes the $\tilde N_1$ Yukawa couplings larger and 
then the washout effect becomes effective. 
This is also suggested by eq.~(\ref{eff}).

In Fig. 2 we show the $L$ asymmetry $|Y_L|$ as a function of $\sin\xi$.  
In the left panel we plot $|Y_L|$ for the various values of $M_1$. 
In the right panel $|Y_L|$ is plotted for the various values of $z_0$.
If we use eqs.~(\ref{eff}) and (\ref{temp}), we find that 
the input parameters adopted to draw Fig.~2 give 
$\tilde m_1\simeq 10^{-3\sim -4}$eV and $T_R>10^7$GeV. 
By using this kind of figures, we can search 
the lower bounds of $M_1$ and $T_R$ required 
to explain the observed $B$ asymmetry for the fixed $M_{2,3}$. 
We practice this analysis changing the values of $M_{2,3}$ within the 
allowed region discussed in the previous part.  
As the result, for the explanation of the $B$ asymmetry based on 
the present model, we find that the lower bounds of $M_1$ and
$T_R$ can be estimated as
\begin{equation}
M_1>5\times 10^8~{\rm GeV}, \qquad
T_R>6\times 10^8~{\rm GeV},
\end{equation} 
and also $\sin\xi$ should be $O(10^{-2})$.

Although the obtained lower bound of the reheating temperature is 
comparable with the lowest value discussed in other neutrino mass 
matrix models where $T_R\simeq 10^{9-10}$GeV is usually suggested, 
we cannot make it much lower. 
Since the lower bound of the mass eigenvalue $\tilde M_1$ is determined 
by eq.~(\ref{mbound}), this result seems not to be avoided in the thermal 
leptogenesis as far as we do not assume the degeneracy among the 
masses of the right-handed neutrinos.
Recently, in \cite{gconst} the upper bound for the reheating 
temperature required from the gravitino problem is estimated as 
$10^{5-7}$GeV if the gravitino has the mass in the range $10^{2-3}$GeV. 
If we do not suppose the light gravitino scenario and we follow this bound, 
the present model is unable to be 
reconciled with the gravitino problem. We need to consider 
the initial $\tilde N_1$ to be yielded in other way. As such a possibility,  
we study the non-thermal leptogenesis in the next part.

\subsection{Nonthermal leptogenesis}
In this subsection we consider that the right-handed neutrinos 
$\tilde N_i$ are produced through the decay of the inflaton. 
This kind of model has been discussed in \cite{inonth}.
The interaction between the inflaton superfield $\Phi$ and 
$\tilde N_i$ is assumed to be given by the superpotential
\begin{equation}
W=\sum_{i=1}^3\lambda_i\Phi\tilde N_i^2.
\end{equation}
After the inflation ends, the inflaton $\phi$ starts to oscillate and 
decays to reheat the universe into the temperature $T_R$.
A part of its oscillation energy $\rho$ of the inflaton is converted into 
$\tilde N_i$ through its decay at $H\simeq \Gamma_\phi$. 
The decay width $\Gamma_\phi$ of the inflaton $\phi$ 
can be expressed as
\begin{equation}
\Gamma_\phi=\sum_{i=1}^3{\lambda_i^2\over 4\pi}m_\phi+\cdots,
\label{width}
\end{equation}
where the ellipses stand for the contribution from other decay modes of the
inflaton and we assume them to be negligible. 
The coupling constants $\lambda_i$ are constrained in such a way as
\begin{equation}
\sum_{i=1}^3\lambda_i^2\simeq 10^{-21}\left({10^{16}~{\rm GeV}\over
m_\phi}\right)\left({T_R\over 10^6~{\rm GeV}}\right)^2, 
\label{coupling}
\end{equation}
which is derived from the condition $H\simeq\Gamma_\phi$.
From this we find that the couplings between the right-handed neutrinos
and the inflaton can be small enough not to affect the inflaton
potential.
The inflaton mass $m_\phi$ can depend on the 
assumed inflation model. However, it should satisfy $m_\phi>\tilde M_i$
to guarantee the inflaton decay into the right-handed neutrinos $\tilde N_i$.

If we use $B_i$ to denote the branching ratio for the decay 
$\phi\rightarrow N_i^2$, we have the energy relation $\rho B_i=\tilde M_in_i$ 
where $\rho={\pi^2\over 30}g_\ast T_R^4$.
Thus the non-thermally generated number density $n_i$ of $\tilde N_i$ can 
be written as
\begin{equation}
{n_i\over s}={3T_RB_i\over 4\tilde M_i}.
\label{init}
\end{equation}
As we mentioned below eq.~(\ref{asym}), $\varepsilon_3$ is much smaller
than $\varepsilon_{1,2}$. In the case dominated by $B_3$ the produced
$L$ asymmetry is expected to be very small unless $T_R$ is rather large.
Then we concentrate our study into the case dominated by $B_{1,2}$. 
Since we find $\varepsilon_1/\tilde M_1=\varepsilon_2/M_2$ from
 eq.~(\ref{asym}), the $L$ asymmetry generated through the immediate decay of
$\tilde N_{1,2}$ is estimated by using  eq.~(\ref{leptasym}) as
\begin{eqnarray}
{n_L\over s}&\simeq& {3\over 16\pi}
{\sqrt{\Delta m_{\rm atm}^2}T_R\over v^2\sin^2\beta}
(\kappa_1 B_1+\kappa_2 B_2)\sin^22\chi \nonumber \\
&\simeq& 10^{-10}\left({T_R\over 10^6~{\rm GeV}}\right)
(\kappa_1B_1+\kappa_2B_2)\sin^22\chi.  
\label{leptasym2}
\end{eqnarray}
Eq.~(\ref{leptasym2}) suggests that $T_R$ should be larger 
than $10^6$~GeV to explain the $B$ asymmetry in any case. 
Moreover, we can find that the $\tilde M_i$ dependence of $n_L/s$
is confined into the washout factor $\kappa_i$. If we consider that
$\tilde M_i$ is larger than $T_R$, the washout effect due to $\tilde
N_i$ is expected to be suppressed by the Boltzmann factor.
Thus for the larger $\tilde M_i$ in such a region,
$n_L/s$ becomes insensitive for the change of the values of $\tilde M_i$.

\begin{figure}[tb]
\begin{center}
\epsfxsize=9cm
\leavevmode
\epsfbox{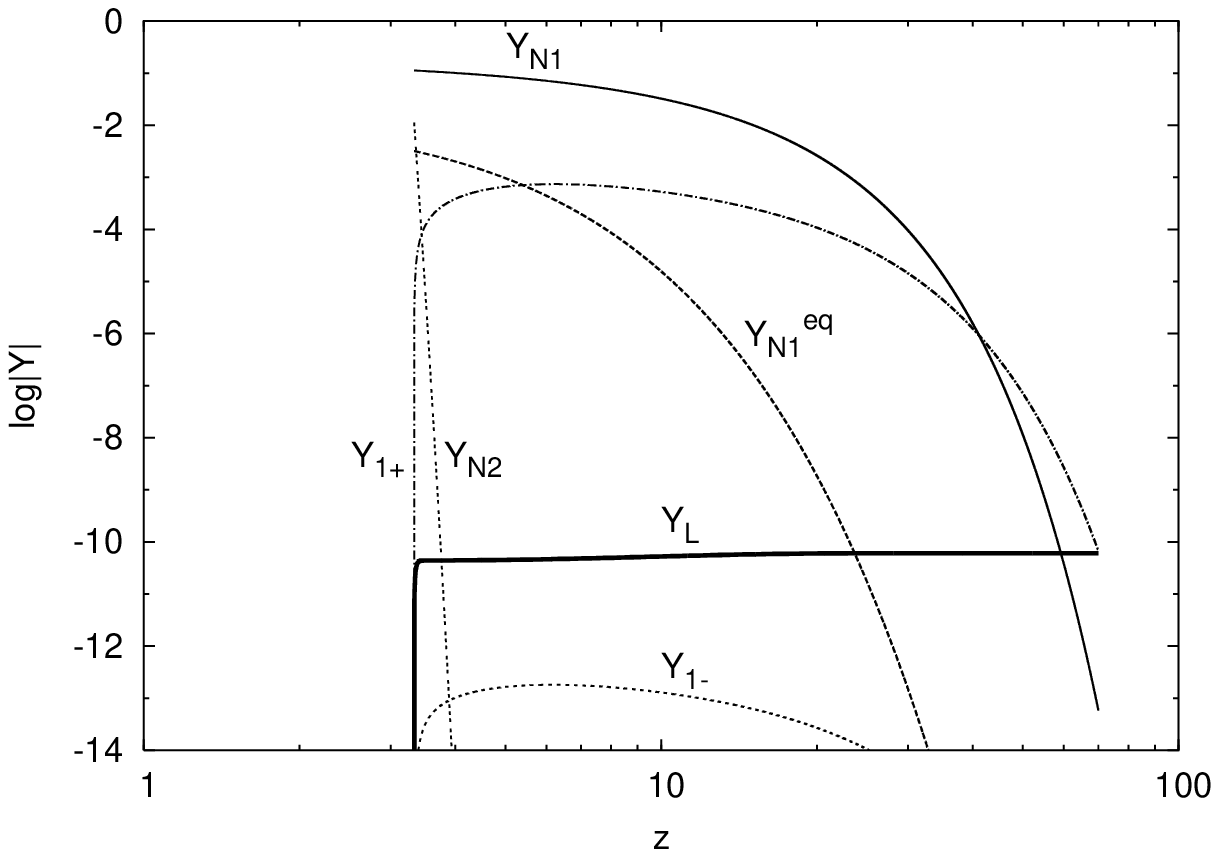}\\
\end{center}
\vspace*{-6mm}
{\footnotesize Fig. 3~~A typical solution of the Boltzmann 
equations in the case of the non-thermal generation of $\tilde N_i$. 
The definitions of $Y_i$ are the same as the ones in Fig.~1. }
\end{figure}

The estimation of the $L$ asymmetry in
eq.~(\ref{leptasym2}) is justified only if $\tilde N_i$ decays into the light
fields immediately after its production \cite{inonth}. 
This requires that $H\simeq\Gamma_\phi~{^<_\sim}~\Gamma_{\tilde N_i}$ should
be satisfied for all $\tilde N_i$ which have the substantial branching ratio
$B_i$.\footnote{In the
construction of the mass matrices presented in appendix A, we find
that $B_2\gg B_1$ is satisfied if the inflaton has no global charges.
However, we consider the general case here.}
Since the inflaton decay width can be estimated by using 
$H\simeq\Gamma_\phi$ as
\begin{equation}
\Gamma_\phi\simeq 0.3g_\ast^{1/2}{T_R^2\over M_{\rm pl}},
\label{infl}
\end{equation} 
we can write the condition for the justification 
of eq.~(\ref{leptasym2}) by applying eqs.~(\ref{coupling}) and 
(\ref{infl}) to $\Gamma_\phi~{^<_\sim}~\Gamma_{\tilde N_1}$ in the form as
\begin{equation}
\left({T_R\over 10^6~{\rm GeV}}\right)^2~{^<_\sim}
~10^6\left({\tilde M_1\over 10^7~{\rm GeV}}\right)
\left({M_2\over 10^{10}~{\rm GeV}}\right)\sin^2\xi.
\label{reheat}
\end{equation}
Since this condition can be easily satisfied for the desirable values of $T_R$,
$\tilde M_1$ and $M_2$, we find that eq.~(\ref{leptasym2}) can be
validated in our interested case. 
However, it is also possible that the immediate decay condition is not
satisfied for $\tilde N_1$. This occurs for 
$\Gamma_{\tilde N_1}<\Gamma_\phi<\Gamma_{\tilde N_2}$ in the 
case of $B_1\simeq B_2$. 
In that case we should take account that the $L$ asymmetry produced
through the
$\tilde N_2$ decay may be washed out by the late entropy release due to the
$\tilde N_1$ decay other than by the usual thermal washout.
This effect is discussed in appendix B. 
We should also check that the condition (\ref{reheat}) can be consistent
with the above mentioned condition $m_\phi>M_{\tilde N_i}$.
This consistency can be easily checked by applying eq.~(\ref{coupling}) 
to $\Gamma_\phi$. 

In order to study the relation between $T_R$ and $\tilde M_i$,
it is useful to classify the situation into three cases:
${\rm (i)}~T_R~{^<_\sim}~\tilde M_1,~  
{\rm (ii)}~\tilde M_1~{^<_\sim}~T_R~{^<_\sim}~M_2,~ 
{\rm (iii)}~M_2~{^<_\sim}~T_R. $
Among these three cases, both the cases (i) and (ii) can easily satisfy 
the condition (\ref{reheat}). On the other hand, the case (iii)
satisfies it only for $M_2~{^<_\sim}~10\tilde M_1\sin^2\xi$, which
requires $\sin\xi\gg 0.1$. 
Thus only the two cases (i) and (ii) seem to be promising 
for the leptogenesis 
consistent with the low reheating temperature.
In fact, $T_R\simeq 10^6$~GeV may be allowed in these cases with 
$B_{1,2}=O(1)$ and $|\sin 2\chi|=O(1)$ if $\kappa_1$ or $\kappa_2$ can
be $O(1)$. 

The washout effect is expected to be mainly caused by the $L$ 
violating interactions due to the thermal $\tilde N_i$.
Since there is the Boltzmann suppression for these processes 
in the case (i), their substantial washout effect cannot be 
expected. On the other hand, in the case (ii) $\tilde N_1$ can
contribute to the washout of the $L$ asymmetry since there is no
large Boltzmann suppression. 
To escape this situation the smallness of the $\tilde N_1$
Yukawa couplings is required. This may be realized for the small 
$\sin\xi$ case. 

To take account of the washout effect quantitatively,
we need to solve the Boltzmann equations numerically by using eq.~(\ref{init})
as the initial value for the $n_i$ at $z_0=M_1/T_R$. 
In Fig. 3 we show a typical solution for the Boltzmann equations in the
case (a). In this figure we assume $|\sin 2\chi|=1$, $B_1=B_2=0.5$ and 
$T_R=3\times 10^6$GeV. 
The input parameters are taken as $M_1=10^7$GeV, 
$M_2=10^8$GeV, $M_3=10^{13}$GeV and $\sin\xi=0.01$. 
This figure shows that the number density of $\tilde N_2$ rapidly
decreases following the Boltzmann distribution.
The $L$ asymmetry reaches the final value faster compared with 
the thermal case.
In the case (b) the sufficient amount of the $L$ asymmetry cannot be
produced. The reason is considered to be the same as the thermal case. 

In Fig. 4 we plot the $L$ asymmetry $|Y_L|$ as the function of $\sin\xi$ 
for various values of the input parameters. 
We calculate $Y_L$ for the typical three 
models with different branching ratios and plot them with the different
symbols, that is, the squares for $B_1=0$, $B_2=1$, the circles for
$B_1=B_2=0.5$ and the triangles for $B_1=1$, $B_2=0$.
The reheating temperature is assumed to be $T_R=3\times 10^6$GeV.
The left panel corresponds to the case (i). In this figure, as the input
parameters we use $M_1=10^8$GeV, $M_2=10^9$GeV, $M_3=10^{13}$GeV for 
three types of the black symbols and 
$M_1=10^8$GeV, $M_2=5\times 10^8$GeV, $M_3=10^{13}$GeV for the white 
symbols. The right panel corresponds to the case (ii). 
In this case, as the input parameters we use $M_1=10^5$GeV, 
$M_2=5\times 10^8$GeV, $M_3=10^{13}$GeV for the black symbols and 
$M_1=10^5$GeV, $M_2=10^8$GeV, $M_3=10^{13}$GeV for the white symbols.
The typical feature in these cases is that the $\sin\xi$ value can be smaller 
compared with the thermal case since we need not to produce $\tilde N_i$
thermally.  

In both panels of Fig.~4 the larger $|Y_L|$ is realized for 
the larger $B_2$ since
the washout effect is smaller compared with the smaller $B_2$ case.
If we make $M_2$ larger keeping $M_1$ fixed in these
figures, $|Y_L|$ becomes a little bit larger but it seems to reach 
almost the upper bound in this setting.
In the left panel the condition (\ref{reheat}) is satisfied 
only for $\sin\xi~{^>_\sim}~10^{-3}$.
In the case of $B_2=1$, however, this condition should be replaced by
$\Gamma_\phi~{^<_\sim}~\Gamma_{\tilde N_2}$ and it is satisfied for all
region of $\sin\xi$ in this figure.
In the case of $B_1=B_2=0.5$, for $\sin\xi<10^{-3}$ we should take 
account of the additional washout effect to the $L$ asymmetry produced
through the $\tilde N_2$ decay, which is
discussed in appendix B. It introduces the
suppression factor $\sqrt{\tilde M_1M_2}\sin\xi/T_R$. It takes a 
value smaller than $O(10^{-1})$ in the present case.
On the other hand, the $L$ asymmetry produced by the late 
$\tilde N_1$ decay cannot be a sufficient amount because of the 
low reheating temperature.
Thus we find that the models with the substantial $B_1$ cannot explain 
the $B$ asymmetry for $\sin\xi<10^{-3}$.
The Yukawa couplings of $\tilde N_1$ and $\tilde N_2$ are proportional to
$\sin\xi$ and $\cos\xi$, respectively.
This fact affects the behavior of $|Y_L|$ as the function $\sin\xi$.
In fact, the figures show the slight increase in the case of $B_2=1$ 
and the decrease in the case of $B_1=1$ when the $\sin\xi$ value becomes
larger.

In the right panel the condition (\ref{reheat}) is satisfied only for 
$\sin\xi>0.1$. The situation is the same as the left panel in the case 
$B_2=1$. Thus in the case (ii) only the model with $B_2\sim 1$ can have the
possibility to explain the $B$ asymmetry.  
The magnitude of $|Y_L|$ becomes smaller for the larger $\sin\xi$ even
in the case of $B_2=1$. Since $\tilde N_1$ can be produced thermally,
the large $\sin\xi$ makes the washout effect more effective,
This feature explains the $|Y_L|$ behavior against $\sin\xi$ 
in the right panel.

We practice this kind of analysis changing the input parameters within
the allowed region. As a result of such study, we find that the 
lower bound of the required 
reheating temperature to explain the $B$ asymmetry is 
$T_R\simeq 3\times 10^6$~GeV for both cases (i) and (ii).

\begin{figure}[tb]
\begin{center}
\epsfxsize=7cm
\leavevmode
\epsfbox{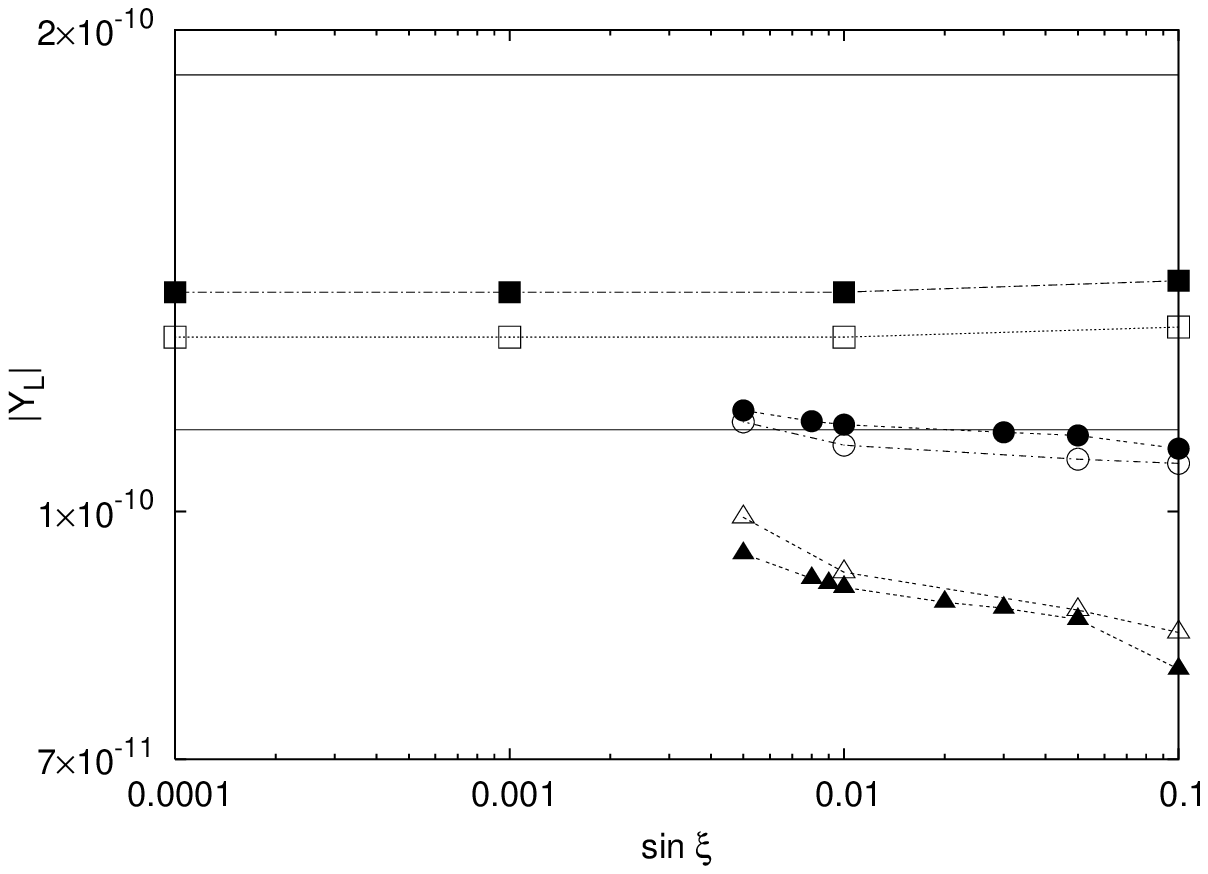}
\hspace*{7mm}
\epsfxsize=7cm
\leavevmode
\epsfbox{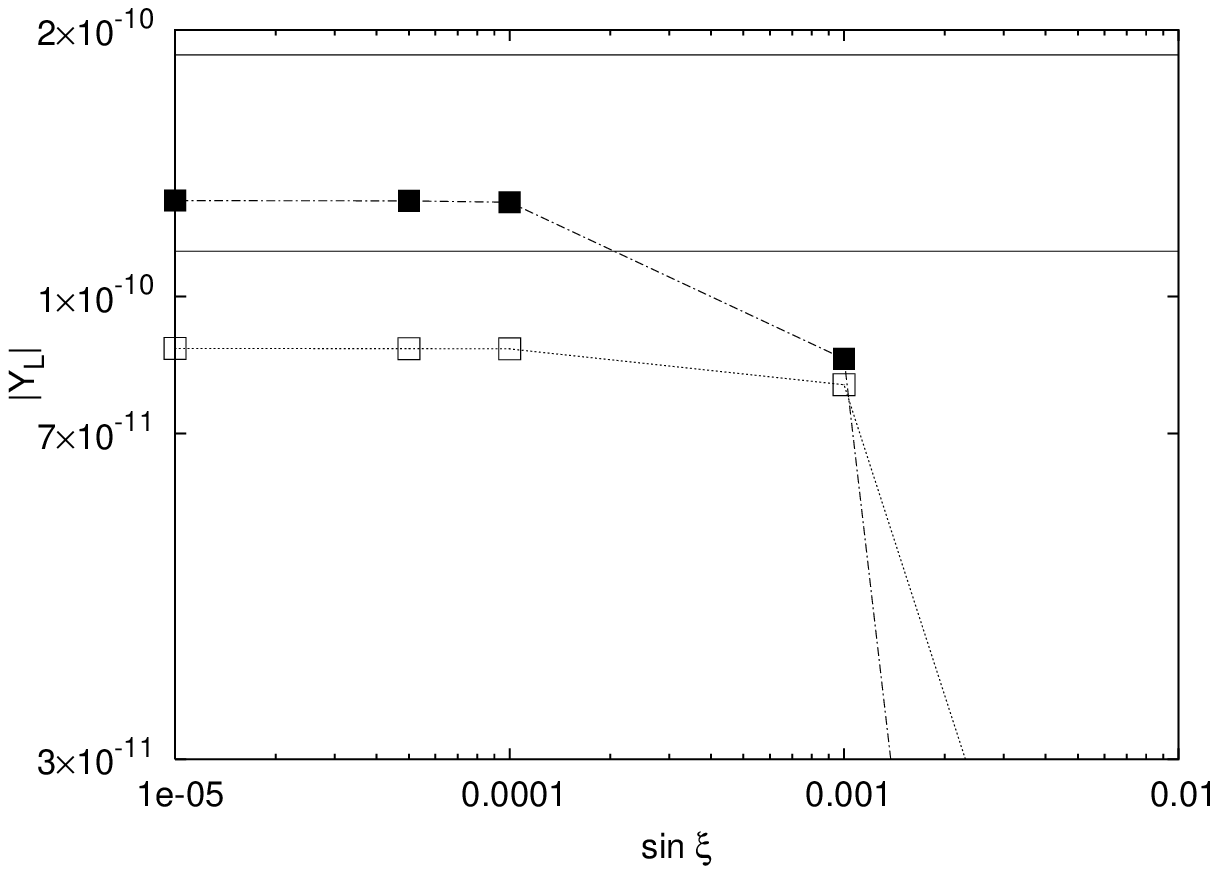}\\
\end{center}
\vspace*{-6mm}
{\footnotesize Fig. 4~~The $L$ asymmetry $Y_L$ as a function of
 $\sin\xi$. Horizontal thin lines represent the desirable region
 to explain the observed $B$ asymmetry. The explanations for the symbols
 are presented in the text.}
\end{figure}

Finally, we briefly comment on other possibility for the non-thermal 
leptogenesis \cite{fnonth}. 
In the early universe the scalar potential of the sneutrino $\tilde N_1$
may be flat enough to deviate largely from its potential 
minimum.\footnote{The sneutrino $\tilde N_1$ can be an inflaton itself
as discussed in \cite{sneut}. However, we do not consider this
possibility  since the reheating temperature is too
high to be reconciled with the gravitino problem in this case.}
If this happens during the inflation, the condensate of $\tilde N_1$
starts to oscillate at $H\simeq \tilde M_1$
and decays at $H\simeq\Gamma_{\tilde N_1}$.\footnote{This oscillation
 may start during the inflation $(\Gamma_\phi<\tilde M_1)$ or 
after the inflation $(\Gamma_\phi>\tilde M_1)$.} 
This oscillation may dominate the energy density of the
universe at a certain time after the reheating due to the inflaton
decay because of its behavior 
as a matter.\footnote{During this oscillation, the
flat direction may store the $L$ asymmetry due to the Affleck-Dine
mechanism as discussed in \cite{fcharge}. 
However, we do not consider this possibility here. }
We assume that it is the case here.

Since its energy density is expressed by 
$\rho_{\tilde N_1}=\tilde M_1^2\vert\tilde N_1\vert^2$, 
the $\tilde N_1$ number density $n_1$ is estimated as 
$\tilde M_1\vert\tilde N_1\vert^2$.
Thus the ratio of the $L$ asymmetry produced 
through its decay to the entropy density is estimated as
\begin{equation}
{n_L \over s}={2\tilde M_1\vert\tilde N_1\vert^2 \over s}\varepsilon_1\kappa_1
={3\over 2}{T_R \over \tilde M_1}\varepsilon_1\kappa_1,
\end{equation}
where in the last equality we use the above mentioned assumption
$\rho_{\tilde N_1}={\pi^2\over 30}g_\ast T^4_R$ for the 
energy density.
If we use eq.~(\ref{asym}) for the CP asymmetry $\varepsilon_1$,
we obtain the similar result for $n_L/s$ to the previous example. 
However, in this case $\Gamma_\phi>\Gamma_{\tilde N_1}$ should be
satisfied and this condition imposes $T_R>\sqrt{10\tilde M_1M_2}\sin\xi$.
Thus the expected $L$ asymmetry is estimated as 
\begin{equation}
{n_L \over s}\simeq 10^{-10}{T_R\over 10^6~{\rm GeV}}~{^>_\sim}~
10^{-10}{\sqrt{\tilde M_1M_2}\sin\xi\over 10^5~{\rm GeV}},
\label{nontherm}
\end{equation}  
where we assume $|\sin 2\chi|=1$ and $\kappa_1=1$.  
This relation gives a constraint for the undetermined parameters 
in the present neutrino mass texture based on the $B$ asymmetry. 
The condition $\Gamma_\phi>\Gamma_{\tilde N_1}$ also requires us
to consider the different type of the inflation from the previous
non-thermal example.

Since $\kappa_1\simeq 1$ is validated only for the case 
$T_R<\tilde M_1$, eq.~(\ref{nontherm}) cannot be applied to the case (b)
in which $T_R~{^>_\sim}~\tilde M_1$ follows.
We can obtain a sufficient amount of the $L$ asymmetry by 
taking $\tilde M_1$ large enough to make $\varepsilon_1$ large 
but keeping $\sin\xi$ small enough to realize $T_R<\tilde M_1$.
Thus, in the case (a) a low reheating temperature like $T_R\simeq 10^6$GeV 
can be enough to produce the required $B$ asymmetry by setting
$M_1$ and $\sin\xi$ suitably. To realize 
such a low reheating temperature, for example, the mass parameters 
in the neutrino sector may be taken as
\begin{equation}
M_1=10^{8}~{\rm GeV}, \quad M_2=10^{10}~{\rm GeV}, \quad 
M_3=10^{13}~{\rm GeV}, \quad \sin\xi=10^{-4}.
\end{equation}
Since the effective mass $\tilde m_1$ is estimated as 
$\tilde m_1\sim 10^{-8}$eV, the washout effect is completely 
negligible as expected. 
We have no gravitino problem in this case since the reheating
temperature realized by the decay of the $\tilde N_1$ condensate 
is sufficiently low comparable to the one given in \cite{gconst}. 

\section{Summary}
We have proposed the neutrino mass matrices in the framework of the
MSSM extended with the three generation right-handed neutrinos.
These mass matrices can realize
the bi-large mixing among the neutrino flavors and explain the neutrino 
oscillation data. 
It can also saturate the upper bound of the CP
asymmetry $\varepsilon_1$ appeared in the leptogenesis. 
Although this model is composed of rather restricted number of parameters, 
it can make the CP asymmetry $\varepsilon_1$ and the effective 
neutrino mass $\tilde m_1$ independent. 
We have applied this model to the thermal and non-thermal 
leptogenesis and studied the influence of
this feature on the reheating temperature, which is crucial for the
cosmological gravitino problem. 

In the thermal leptogenesis our neutrino mass texture seems not to
be able to  make the reheating temperature required from the 
explanation of the $B$ asymmetry low enough to be 
consistent with the gravitino problem.
However, it seems to be able to realize a value near the lower bound of the 
reheating temperature obtained in the thermal leptogenesis framework.
In the non-thermal case we have found that the low reheating temperature 
consistent with the gravitino problem can be sufficient for the successful 
leptogenesis. Even in that case the parameters in the neutrino
mass matrices can be consistent with the neutrino oscillation data.

As is shown in this study, some kinds of neutrino mass texture can be
constrained by the leptogenesis. 
It may be worthy to proceed a lot of study based on the concrete
neutrino model to clarify the relation among the neutrino mass texture, 
the leptogenesis and the gravitino problem.  
\vspace{.5cm}

This work is partially supported by a Grant-in-Aid for Scientific
Research (C) from Japan Society for Promotion of Science (No.14540251).
\newpage
\noindent
{\Large\bf Appendix A}

In this appendix we present an example of the construction for
the assumed texture of the neutrino mass matrix.
We consider the Frogatt-Nielsen type global flavor symmetry U(1)$^5$
and an additional discrete $Z_2$ symmetry in the lepton sector. 
The charge assignment of the chiral superfields
for the symmetry U(1)$^5\times Z_2$ is assumed as
\begin{eqnarray}
&&N_1~(1,1,1,0,1;+), \quad N_2~(1,0,1,1,1;+),  \quad N_3~(1,1,0,1,0;+),
\nonumber\\
&&L_1~(0,0,-1,-1,0;+), \quad L_2~(-1,0,0,-1,0;+), \quad L_3~(0,-1,0,-1,0;+).
\end{eqnarray}
The Higgs chiral superfield $H_2$ is neutral for this symmetry.
In order to realize the hierarchical structure of the mass matrices,
we introduce the following several chiral superfields which are 
singlet for the standard gauge groups:
\begin{equation}
\begin{array}{ccc}
\phi_1~(-1,-1,-1,0,0;-) &\phi_2~(-1,0,-1,-1,0;-), & 
\phi_3~(-2,-2,0,-2,0;+),\\
\chi_1~(-1,0,0,0,0;+), & \chi_2~(0,-1,0,0,0;+), & 
\chi_3~(0,0,-1,0,0;+), \\
&\eta~(0,0,0,0,-1;+).& \\
\end{array}
\end{equation}
   
If we assume that the scalar components of these superfields get vacuum
expectation values defined by
\begin{equation}
\epsilon_i\equiv {\langle\phi_i\rangle\over M_{\rm pl}}, \qquad
\delta\equiv {\langle\chi_i\rangle\over M_{\rm pl}}, \qquad
\zeta\equiv {\langle\eta\rangle\over M_{\rm pl}},
\end{equation}
we can obtain both the right-handed Majorana mass matrix 
and the Dirac mass matrix as follows:
\begin{equation}
{\cal M}\simeq M_3\left(\begin{array}{ccc}
\epsilon_1^2\zeta^2/\epsilon_3 & \epsilon_1\epsilon_2\zeta^2/\epsilon_3
 & 0 \\
\epsilon_1\epsilon_2\zeta^2/\epsilon_3 & \epsilon_2^2\zeta^2/\epsilon_3
 & 0 \\
0 & 0 & 1 \\ \end{array}\right), \qquad 
m_D=v_2\left(\begin{array}{ccc}
0 & 0&0\\ \zeta\delta &\zeta\delta & 0 \\ 
0 &\delta & \delta \\
\end{array}\right),
\label{const}
\end{equation}
where $M_3\equiv M_{\rm pl}\epsilon_3$ and the order one 
coefficients are abbreviated.  The difference between the cases
(a) and (b) should be considered to be explained by these coefficients.

We can check that these mass matrices can consistently realize the
texture assumed in the text.
Comparing $m_D$ in eq.~(\ref{mmatrix}) with that in eq.~(\ref{const}) and 
also using eq.~(\ref{nmass}), we find 
\begin{equation}
\delta\sim 0.1 \left({M_3 \over 10^{13}{\rm GeV}}\right)^{1/2}, \qquad 
\zeta\sim  0.4\left({M_2\over M_3}\right)^{1/2}, \qquad
\epsilon_3={M_3\over M_{\rm pl}}.
\end{equation}
Applying this result to ${\cal M}$ in eq.~(\ref{mmatrix}) and 
eq.~(\ref{const}), we obtain
\begin{equation}
M_1\sim \left({\epsilon_1\over\epsilon_2}\right)^2 M_2, \qquad
\epsilon_3\sim 0.16\epsilon_2^2, \qquad  
\sin\xi\sim {\epsilon_1\over\epsilon_2}.
\end{equation}
If we take $\delta\sim 0.1$, $\zeta\sim 0.03$ and 
$\epsilon_2\simeq 30\epsilon_1$, for example, we find 
\begin{equation}
M_1\sim 10^8~{\rm GeV},\quad 
M_2\sim 10^{10}~{\rm GeV},\quad M_3\sim 10^{13}~{\rm GeV},\quad 
\sin\xi\sim 0.03.
\end{equation}
This suggests that (\ref{const}) can realize the mass matrices assumed
in the text. 

We can consider the couplings of the right-handed neutrinos $\tilde N_i$
to the inflaton $\phi$, which is assumed to be neutral under the present
global symmetry. The lowest order superpotential allowed by this
symmetry is written as
\begin{equation}
W=c_1{\phi_1^2\eta^2\over M_{\rm pl}^4}\phi N_1^2+
c_2{\phi_2^2\eta^2\over M_{\rm pl}^4}\phi N_2^2
+c_3{\phi_3\over M_{\rm pl}}\phi N_3^2,
\end{equation}
where the coefficients $c_{1,2,3}$ are assumed to be O(1).
Thus the coupling constants $\lambda_i$ are estimated as
\begin{equation}
\lambda_1=c_1\epsilon_1^2\zeta^2, \qquad \lambda_2=
c_2\epsilon_2^2\zeta^2, \qquad \lambda_3=c_3\epsilon_3. 
\end{equation}
We find that these couplings can be consistent with the 
low value of $T_R$ if $m_\phi\simeq 10^{12}$GeV is assumed.
In this case $B_2\gg B_1$ can be satisfied since $B_3=0$ is realized for
$M_3\simeq 10^{13}$GeV.  
\vspace*{10mm}

\noindent
{\Large\bf Appendix B}

In the non-thermal leptogenesis the $L$ asymmetry produced through the
$\tilde N_2$ 
decay may be washed out in a different way compared with the thermal
case. If $\Gamma_{\tilde N_1}<\Gamma_\phi{^<_\sim}~\Gamma_{\tilde N_2}$ 
is satisfied, $\tilde N_2$ decays into the
light particles at the time $t_2$ immediately after the inflaton 
decays into $\tilde N_2$.
The decay products of $\tilde N_2$
behaves as the radiation and its energy density decreases as 
$\rho_{\tilde N_2}\propto a^{-4}$
where $a$ is the cosmological scale parameter.
On the other hand, $\tilde N_1$ decays at the time 
$t_1$ after the completion of the
inflaton decay. If $\tilde N_1$ behaves as a matter because 
of $T_R<\tilde M_1$, its energy density decreases as 
$\rho_\phi\propto a^{-3}$.  
Then the cosmological energy density may be dominated by the 
$\tilde N_1$ energy at least as far as $B_1$ and $ B_2$ are the same order.
The additional washout can occur in such a case.

The cosmological energy density $\rho(t_2)$ and the temperature
$T_2(t_2)$ of its decay products can be expressed as
\begin{equation}
\rho(t_2)B_2={\pi^2\over 30}g_\ast(t_2)T_2^4(t_2), \quad
H^2(t_2)={\rho(t_2)\over 3M_{\rm pl}^2}\simeq\Gamma_\phi^2.
\end{equation}
Taking account of these, we can find that the temperature $T_2$ of the
decay products of $\tilde N_2$ satisfies the relation such as
\begin{equation}
\left({T_2(t_2)\over T_2(t_1)}\right)^4=
\left({a(t_1)\over a(t_2)}\right)^4
=\left({H(t_2)\over H(t_1)}\right)^{8/3}
=\left({\Gamma_\phi\over \Gamma_{\tilde N_1}}\right)^{8/3}.
\end{equation}
From this relation we obtain
\begin{equation}
T_2(t_2)=T_2(t_1)\left({ \Gamma_\phi\over\Gamma_{\tilde N_1}}\right)^{2/3}.
\end{equation}

Now we can estimate the entropy production of the late decay of $\tilde
N_1$. Since $H^2(t_1)=\rho(t_1)/3M_{\rm pl}^2=\Gamma_{\tilde N_1}$ 
is satisfied,
the ratio between the entropy density $s_b(t_1)$ before the $\tilde N_1$ decay
and the entropy density $s_a(t_1)$ after the decay can be written as
\begin{equation}
{s_b(t_1)\over s_a(t_1)}={g_\ast(t_2)T_2^3(t_1)\over g_\ast(t_1)T_1^3(t_1)}
\simeq\left({T_2(t_2) \over T_1(t_1)}\right)^3
\left({\Gamma_{\tilde N_1}\over\Gamma_\phi}\right)^2
\simeq \left({\Gamma_{\tilde N_1} \over \Gamma_\phi}\right)^{1/2}.
\end{equation} 
By using the expression of each decay width, we find that $\kappa_2$ 
is written as
\begin{equation}
\kappa_2\simeq\kappa {\sqrt{\tilde M_1M_2}\over T_R}\sin\xi,
\end{equation}
where $\kappa$ is the usual thermal washout effect due to the $L$ violating 
scattering mediated by the right-handed neutrinos and so on. 


\end{document}